\documentstyle[aps,prl,psfig,preprint]{revtex}

\newcommand{\lcco}{$\rm La_{14-x}Ca_xCu_{24}O_{41}$}

\newcommand\dh{$\Delta H$}
\newcommand\dht{$\Delta H(T)$}
\newcommand\cu{$\rm Cu^{2+}$}

\begin{document}

\draft

\title{Strong anisotropy of superexchange
in the copper-oxygen chains of $\bf La_{14-x}Ca_xCu_{24}O_{41}$ }

\author{V. Kataev\cite{address}, K.-Y. Choi, M. Gr\"{u}ninger,
U. Ammerahl, B. B\"{u}chner, A. Freimuth  and A. Revcolevschi$^{\dag}$}

\address{II. Physikalisches Institut, Universit\"{a}t zu K\"{o}ln,
50937 K\"{o}ln, Germany\\
$^{\dag}$ Laboratoire de Chimie des Solides,
Universit\'e Paris-Sud, 91405 Orsay C\'edex, France}

\date{November 16, 2000}

\maketitle

\begin{abstract}
Electron spin resonance data of \cu\ ions in \lcco\ crystals
$(x=9,11,12)$ reveal a very large width of the resonance line in
the paramagnetic state. This signals an unusually strong
anisotropy of $\sim \! 10\%$ of the isotropic Heisenberg
superexchange in the Cu-O chains of this compound. The strong
anisotropy can be explained by the specific geometry of two
symmetrical 90$^\circ$ Cu-O-Cu bonds, which boosts the importance
of orbital degrees of freedom. Our data show the apparent
limitations of the applicability of an isotropic Heisenberg model
to the low dimensional cuprates.
\end{abstract}



Magnetic interactions in strongly correlated transition metal
oxides have attracted much interest during the past decade due to
their intimate relationship with high temperature
superconductivity in the cuprates, but also in connection with
novel quantum magnetic phenomena in spin-chain and spin-ladder
materials. The superexchange interaction between magnetic ions in
oxides is mediated via oxygen ligands and is described by the
Hamiltonian
\begin{equation}
{\cal H}=J_{\rm iso}\sum_{ij}{\bf S}_i{\bf S}_j + \sum_{ij}{\bf
d}_{ij}[{\bf S}_i\times{\bf S}_j] + \sum_{ij}{\bf S}_i A_{ij}
{\bf S}_j \; ,
\label{hamilton}
\end{equation}
where the first term denotes the isotropic Heisenberg exchange,
and the second and third represent antisymmetric and symmetric
{\it anisotropic} contributions caused by spin-orbit coupling. In
the case of a \cu\ ion with a single hole with $S$=1/2 in the 3$d$
shell, the orbital angular momentum is quenched in the solid due
to crystal field splitting. In general spin-orbit coupling and the
corresponding anisotropies play only a minor role in cuprates. For
instance the $\rm CuO_2$ planes of the parent compounds of high
temperature superconductors like $\rm La_2CuO_4$ or
Sr$_2$CuO$_2$Cl$_2$ are thought to be the best representatives of
a 2D $S$=1/2 square lattice {\em isotropic} Heisenberg
antiferromagnet with $J_{\rm iso}\!\approx \! 1500$ K.\@ This
large value of $J_{\rm iso}$ is typical for the 180$^\circ$
Cu-O-Cu bond angle present in 2D cuprates. The anisotropic
magnetic couplings $A_{ij}\ll \mid\!{\bf d}_{ij}\!\mid$ both do
not exceed 1\% of $J_{\rm iso}$ in $\rm La_2CuO_4$ \cite{Kastner}.
Nevertheless, this small anisotropy determines the orientation of
spins with respect to the lattice in the magnetically ordered
state and is responsible for such remarkable phenomena as weak
ferromagnetism or the presence of spin wave gaps.

\begin{figure}
\centerline{\psfig{figure=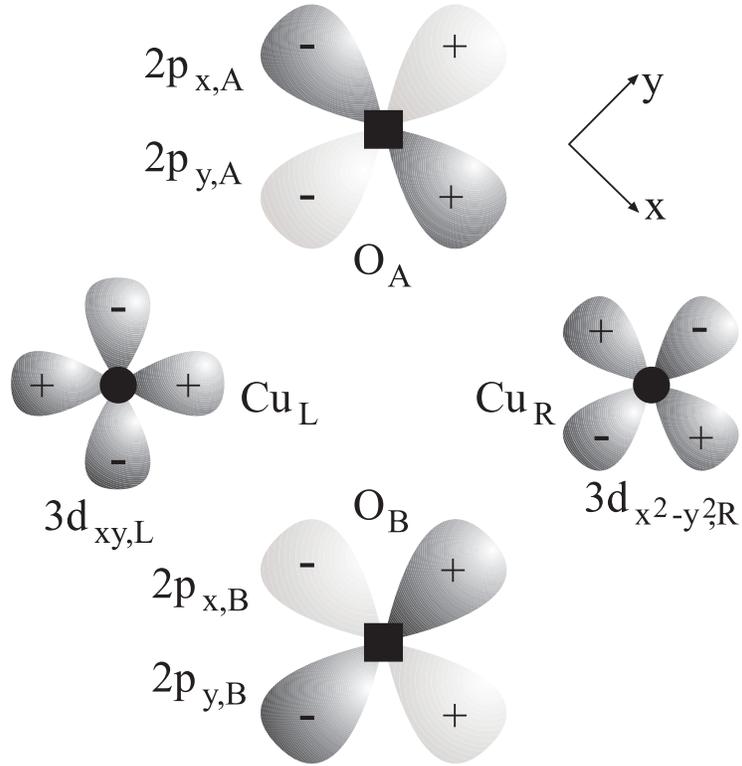,width=0.6\columnwidth,angle=0,clip=}}
\caption{Copper and oxygen orbitals of two symmetric 90$^\circ$
Cu-O-Cu bonds relevant for the anisotropic coupling.}
\label{orbitals}
\end{figure}

According to the Goodenough-Kanamori-Anderson rules \cite{GKA} the
strength of the leading isotropic coupling $J_{\rm iso}$ can be
considerably reduced by decreasing the Cu-O-Cu bond angle from
180$^\circ$ to 90$^\circ$. In the case of a 180$^\circ$ bond the
antiferromagnetic (AF) coupling is mediated via a single ligand
orbital, whereas the exchange in a 90$^\circ$ bond proceeds via
orthogonal orbitals (see Fig. \ref{orbitals}) which are coupled
via Hund's rule, resulting in a {\em ferromagnetic} coupling
\cite{khomskii}. Experimentally, one still finds an AF $J_{\rm
iso}\!\approx\!$ 120 K \cite{cugeo} in the quasi-one-dimensional
Cu-O chains of the spin-Peierls compound $\rm CuGeO_3$ with a bond
angle $\sim\! 98^\circ$. Still closer to 90$^\circ$ $J_{\rm
iso}$ indeed changes sign and becomes ferromagnetic, as in
La$_{14-x}$Ca$_x$Cu$_{24}$O$_{41}$ \cite{Carter}. The dependence
of the {\em anisotropic} corrections on the bond angle have not
been studied to the same extent. Recently, it has been predicted
theoretically that the magnetic anisotropy of two symmetrical
90$^\circ$ Cu-O-Cu bonds may be unusually large
\cite{Yushan,Aharony}. This bonding geometry arises if Cu-O
plaquettes (i.e.\ four oxygens forming a square with a Cu ion at
the center) are connected along their edges and is found in
several systems like La$_{14-x}$Ca$_x$Cu$_{24}$O$_{41}$
\cite{structure}, Li$_2$CuO$_2$ \cite{Hayashi} or $\rm
Ca_2Y_2Cu_5O_{10}$ \cite{fong}.

Electron spin resonance (ESR) is a very sensitive tool to study
the anisotropy of the spin-spin coupling which, in particular, is
the main source of the finite width \dh\ of an ESR signal in
concentrated paramagnets \cite{Abragam}. In this paper we present
ESR data of \lcco\ (LCCO) single crystals, which show a very broad
\cu\ ESR signal of $\Delta H\! \sim\! 1500\! -\! 2000$ Oe from the
Cu-O chains in the paramagnetic regime at $T>60$ K. Our analysis
shows that the observed width of the ESR signal requires an
unusually large anisotropic coupling of about 10\% of the leading
isotropic exchange $J_{\rm  iso}$. This cannot be explained by
conventional estimates of the dipole-dipole or anisotropic
exchange interactions which neglect the geometry. The data thus
give evidence for a significantly enhanced contribution of
spin-orbit coupling to the magnetism of copper oxides in certain
bonding geometries.

ESR measurements were carried out at a frequency of 9.47 GHz. The
magnetic susceptibility $\chi_{\rm static}$ was measured with a
SQUID-magnetometer at a field of 1 Tesla. \lcco\ single crystals
with $x$=9, 11 and 12 were grown by the travelling solvent
floating zone method \cite{Udo}. Their single phase structure and
stoichiometry have been verified by x-ray, energy dispersive x-ray
and thermogravimetric analyses. The Ca content determines the
average oxidation state of $\rm Cu^{2+\delta}$. The average number
of holes per Cu site increases from $\delta \! \approx \! 0.04$
for $x$=9 to $\delta \! \approx \! 0.17$ for $x$=12. The chains of
edge-sharing Cu-O plaquettes are parallel to the $c$-axis and lie
in one crystallographic plane ($ac$-plane). The exchange
interaction in the chains is found to be small and ferromagnetic
($J_{\rm iso}\!\approx\! -20$ K \cite{Carter}), as expected for
the nearly 90$^\circ$ Cu-O-Cu bond angle. Additionally, \lcco\
contains planes of $\rm Cu_2O_3$ two-leg spin ladders running
parallel to the chains \cite{structure}. The ladders show a
spin-singlet ground state with a spin gap of the order of 500-600
K \cite{Kumagai,Imai}.

\begin{figure}
\centerline{\psfig{figure=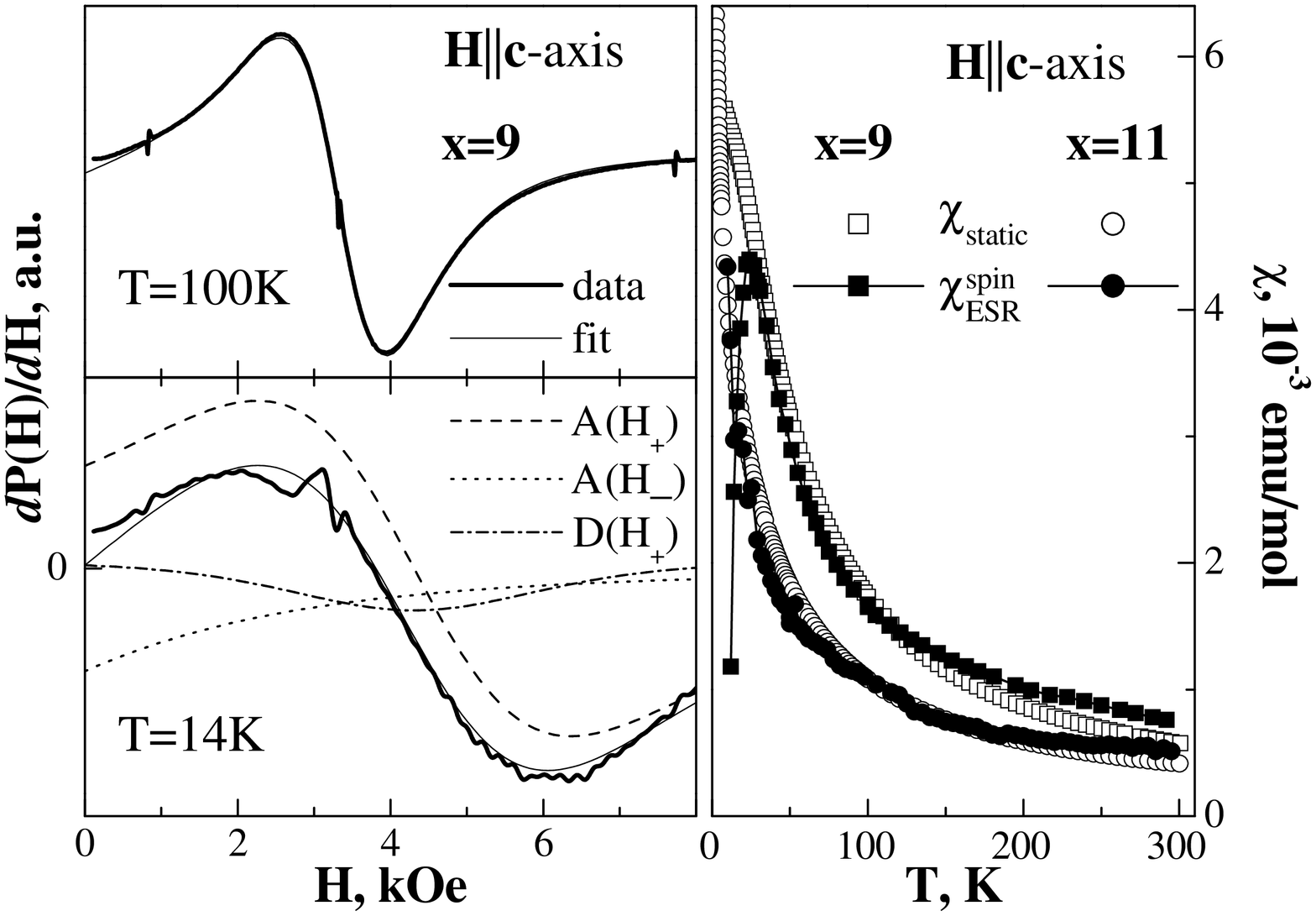,width=0.95\columnwidth,angle=-90,clip=}}
\caption{Left: Representative ESR spectra (derivatives of the
absorbed microwave power $dP(H)/dH$, thick lines) of a \lcco\
crystal ($x$=9) at $T$=100 and 14 K and the respective fits
according to Eq.\ \protect\ref{fit} (thin solid lines). Different
contributions to the $T$=14 K fitting curve  are shown by broken
lines. Right: A comparison of the spin susceptibility $\chi_{\rm
ESR}^{\rm spin}$ derived from the ESR intensity $I(T)$ with the
static susceptibility $\chi_{\rm static}$ for $x$=9 and 11.}
\label{spectrum}
\end{figure}

\begin{figure}
\centerline{\psfig{figure=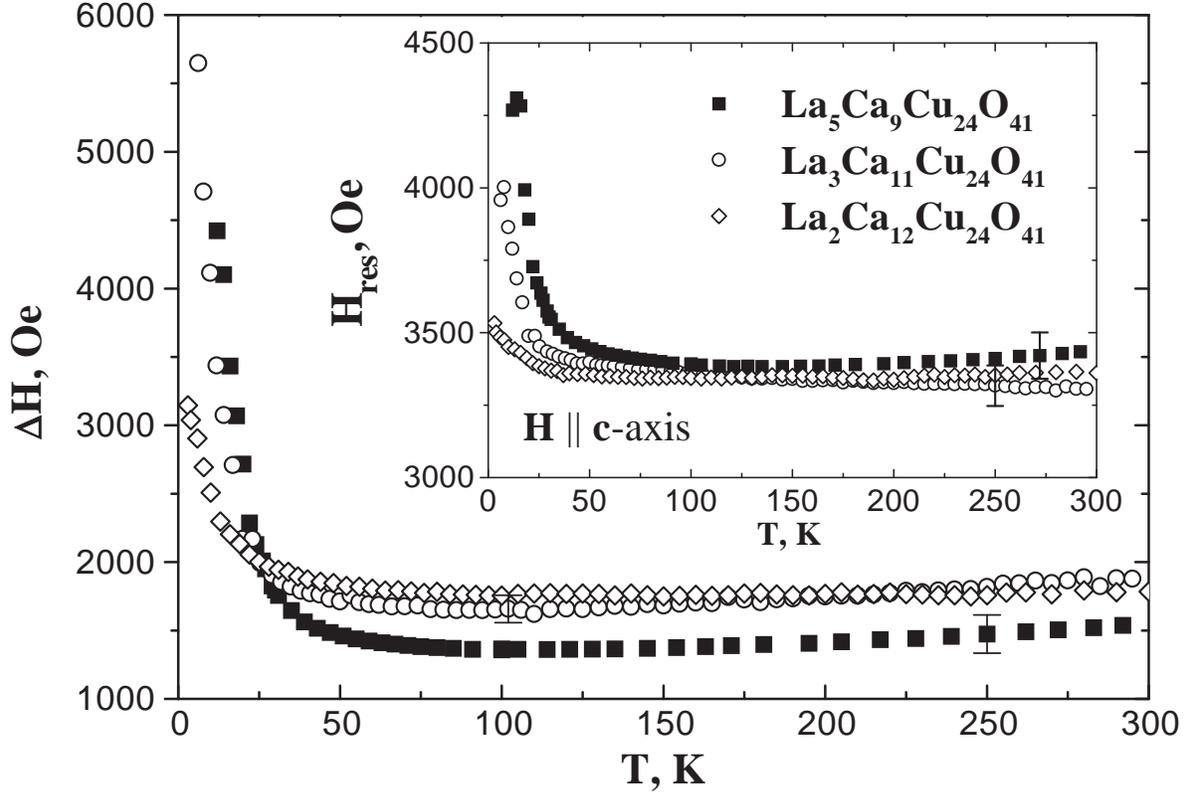,width=0.95\columnwidth,angle=-90,clip=}}
\caption{Temperature dependence of the ESR line width \dht\ and of
the resonance field $H_{\rm res}(T)$ (inset) for \lcco\ crystals
with $x$=9, 11 and 12 ($H\parallel c$).} \label{dH.Hres}
\end{figure}

Representative ESR spectra of the sample with $x$=9 and the fitting curves
are shown in the left panels of Fig.\ \ref{spectrum} at $T=100$ and 14 K.\@
Since the width of the signal \dh\ is comparable to the value
of the resonance field $H_{\rm res}$, the fitting function $f(H)$
has to include Lorentzian absorption derivatives corresponding to
both, right and left circularly polarized components $A(H_+)$ and
$A(H_-)$ of the linearly polarized microwave field \cite{Abragam}.
Moreover, we add to $f(H)$ the correction due to the non-diagonal
contributions to the dynamic susceptibility which appears as an
admixture of the Lorentzian dispersion $D(H_+)$ \cite{Benner1}:
\begin{eqnarray}
f(H)=A(H_+) + A(H_-) + D(H_+) \; ,
\label{fit}
\\
A(H_\pm )=-\frac{16ah_\pm }{(3\!+\!h_\pm^2)^2} \; , \;\;
D(H_\pm )=\frac{3d(3\!-\!h_\pm^2)}{(3\!+\!h_\pm^2)^2} \;  .
\nonumber
\end{eqnarray}
Here, $h_{\pm}\!=\!2(H\mp H_{res})/\Delta H$,
and $a$ and $d$ $(< a)$ are the amplitudes of the Lorentzian
absorption and dispersion signals, respectively. From this fit we
obtain the values of $H_{\rm res}$,  \dh\ and the integrated
intensity $I$ of the ESR line. The temperature dependence of these
quantities for $H\parallel c$ \cite{why.for.c} is shown in Figs.\
\ref{spectrum} (right panel) and \ref{dH.Hres}. From the dependence of
$H_{\rm res}$ on the orientation of the magnetic field we derive
$g$-factors of $g_c=2.02\pm 0.02$ and $g_b=2.30\pm 0.05$
\cite{comment}, which are typical for \cu\ ions in a four-fold
square oxygen coordination \cite{Abragam}.

In the {\em paramagnetic} regime, the ESR intensity $I(T)$ is per
definition proportional to the static susceptibility of the
resonating spins $\chi^{\rm spin}_{\rm  ESR}$ \cite{Abragam}. In
order to obtain $\chi^{\rm spin}_{\rm  ESR}$ in absolute units we
have calibrated $I(T)$ of the samples at $T=100$ K against the
intensity of the simultaneously measured spectrum of the standard
reference material $\rm Al_2O_3 + 0.03\% Cr^{3+}$ (the latter
spectrum is visible in the top left panel of Fig.\ \ref{spectrum}
as three small background lines at 840, 3340 and 7740 Oe). The $T$
dependences of $\chi^{\rm spin}_{\rm  ESR}$ and $\chi_{\rm
static}$ are very similar for $T>30$ K  (right panel of Fig.\
\ref{spectrum}). The deviations at $T\lesssim 30$ K are related to
a transition to a long range ordered AF state (see below). For
$50\ {\rm K}<T<300$ K, $\chi_{\rm static}$ in LCCO is entirely
described in terms of weakly interacting Cu-O spin chains, the
contribution from the ladders is negligible due to their large
spin gap \cite{Carter}. From the similarity between $\chi^{\rm
spin}_{\rm  ESR}$ and $\chi_{\rm  static}$ we therefore conclude
that the ESR spectrum can be ascribed solely to the Cu spins in
the chains.

The ESR line width in magnetic insulators is usually decomposed
into non-critical and critical parts \cite{Huber}:
\begin{equation}
\Delta H=\Delta H(\infty) + \Delta H_{\rm crit}(T).
\label{totalwidth}
\end{equation}
The first term in Eq.\ \ref{totalwidth} is determined by spin-spin
interactions at high temperatures, where spins are uncorrelated
({\em paramagnetic} regime). The second term $\Delta H_{\rm crit}(T)$ defines
an additional contribution to the line width which arises at low temperatures,
when long-range magnetic order is approached. It is caused by the fluctuations
of the staggered magnetization in the short-range ordered state. These
enhance the rate of the spin-spin relaxation $1/T_2$ and
consequently broaden the ESR line additionally \cite{Huber}.
According to the analysis of $\chi_{\rm  static}$ in Ref.\
\cite{Carter}, insulating LCCO is paramagnetic above 50 K, whereas
at lower temperatures spin correlations grow and finally result in
long range AF order, which has been observed in $\rm
La_5Ca_9Cu_{24}O_{41}$ at $T_N\!\approx \!$ 10 K
\cite{Ammerahl,Matsuda4}. Indeed, both the considerable shift of
$H_{\rm res}$ to higher fields and the broadening of the ESR line
at $T\!<\!$ 60 K (see Fig.\ \ref{dH.Hres}) are obviously due to the
development of short-range magnetic order ($\Delta H_{\rm crit}(T)$
in Eq.\ \ref{totalwidth}) \cite{Benner2}. The intensity of the ESR
signal drops rapidly already above $T_N$ due to the growth of
short-range AF ordered regions. Most of the spins within these
regions do not contribute to the ESR spectrum since their resonance
frequency is outside the range of our spectrometer. This explains
the discrepancy between $\chi^{\rm spin}_{\rm  ESR}$ and $\chi_{\rm
static}$ at low $T$ (right panel of Fig.\ \ref{spectrum}). With
increasing Ca content the critical behavior of the ESR response
gets less pronounced thus giving evidence for a rapid suppression
of AF correlations in LCCO by hole doping.

The main concern of the present paper is the large value of \dht\
of about 1500-2000 Oe for temperatures far away from the
magnetically ordered state (see Fig. \ref{dH.Hres}). It is
reasonable to assume that the almost $T$-independent ESR line
width in the paramagnetic regime above 60-80 K is determined by
the Cu spin-spin interactions, i.e. by $\Delta H(\infty)$ in
Eq.\ref{totalwidth} \cite{inhomogen}. Magnetic resonance in
paramagnetic insulators is usually discussed in terms of the
``moments'' of a Gaussian absorption line \cite{Abragam}. In
particular, $\Delta H(\infty)$ is proportional to the second
moment $M_2$, which is determined only by various {\it
anisotropic} interactions between spins. On the other hand, the
isotropic Heisenberg exchange interaction $J_{\rm iso}\sum {\bf
S}_i{\bf S}_j$ influences the shape of the line via the well known
effect of ``exchange narrowing'' of the magnetic resonance which
takes place if $\mid \!\! J_{\rm iso}\!\! \mid \! /g\mu_B\gg
H_{\rm res}\gtrsim\Delta H$. In this case the ESR signal acquires
a Lorentzian line shape, which is indeed observed in LCCO (see
Fig.\ \ref{spectrum}). The corresponding ``exchange narrowed''
line width reads \cite{Kubo,dimension}:
\begin{equation}
\Delta H(\infty)\simeq\frac{\hbar^2}{g\mu_B}\frac{M_2}{\mid\!\!
J_{\rm iso}\!\!\mid}. \label{exchnarrow}
\end{equation}

The most obvious anisotropic interaction which broadens the
resonance line is a dipole-dipole interaction between the spins
${\bf S}_i$ and ${\bf S}_j$ separated by a distance ${\bf
r}_{ij}$: ${\cal H}_{\rm  dd}=g^2\mu_B^2\sum_{ij}r^{-3}_{ij}[{\bf
S}_i{\bf S}_j-3r^{-2}_{ij} ({\bf S}_i{\bf r}_{ij})({\bf S}_j{\bf
r}_{ij})]$. This yields an exchange narrowed dipole-dipole
contribution to the width of $\Delta H^{\rm dd}\!\approx \! 1$ Oe,
which is negligible compared to the experimental result.

Now we focus on the contributions to \dh\ arising from the {\em
anisotropy} of exchange (last two terms in Eq.\ \ref{hamilton}).
The term $\sum_{ij}{\bf d}_{ij}[{\bf S}_i\times{\bf S}_j]$ is
known as the antisymmetric Dzyaloshinsky-Moriya (DM) interaction
\cite{Moriya,Dzy}. However, in LCCO the DM vector ${\bf d}_{ij}$
is zero due to the presence of an inversion center between two
nearest neighbor Cu sites \cite{structure}. The only possible
source of the large \dh\ in LCCO is hence the symmetric
anisotropic exchange $\sum_{ij}{\bf S}_i A_{ij} {\bf S}_j$. In
this case $M_2 \!\simeq \! A^2_{ij}$ \cite{Abragam}. With the
observed value of $\Delta H \!\approx\! 1500$ Oe and $J_{\rm
iso}\!\approx \!$ 20 K \cite{Carter} we get from Eq.
\ref{exchnarrow} a surprisingly strong anisotropy $A_{ij}$ of
about 10\% of $J_{\rm iso}$. This value of $A_{ij}$ is ten times
larger than the conventional estimate \cite{Moriya} $A^{\rm
conv}_{ij} \simeq (\Delta g/g)^2J_{\rm iso}\approx$ 1\% of $J_{\rm
iso}$, where $\Delta g\approx 0.2$ is the average deviation of the
$g$-factor of \cu\ in LCCO from its spin-only value $g$=2. In
terms of experimentally accessible quantities, the conventional
estimate yields $\Delta H\approx 15$ Oe, whereas the experimental
result is two orders of magnitude larger.

In the following we discuss the basic physics that may cause such
an unusually strong anisotropy of superexchange in
LCCO\cite{Yushan,Aharony}. For a qualitative understanding it is
sufficient to consider only two nearest neighbor Cu ions, Cu$_L$
and Cu$_R$ in the Cu-O chain (see Fig.\ \ref{orbitals}). In
perturbation theory, the superexchange interaction arises from
virtual hopping processes via the two intermediate oxygen ligands
(A and B), which couple the Cu spins by 90$^\circ$ bonds. In hole
notation oxygen orbitals are empty. The crystal field splits the
$3d^9$ state of a Cu$^{2+}$ ion, and the lowest orbital
$d_{x^2-y^2}$ is non-degenerate and singly occupied. Only the
twofold Kramers spin degeneracy remains in the ground state.
However, a finite spin-orbit coupling $\lambda {\bf
L}\!\cdot\!{\bf S}$ couples the $|x^2\!-\!y^2\!\!>$ state with the
orbital states $|xy\!\!>$, $|yz\!\!>$ and $|zx\!\!>$. The central
point is that the relevance of these orbitals for superexchange
depends on the geometry of the considered bond. In case of a
180$^\circ$ bond directed e.g. along the $x$-axis with a single
oxygen ligand at the midpoint between the two Cu sites, the
symmetry and the contribution of the $|xy,{\rm Cu_L}\!\!>$ and
$|zx,{\rm Cu_L}\!\!>$ states are {\em identical} with respect to
the Cu-O-Cu bond. In the present case, on the other hand, exchange
processes involving virtual hopping of the hole from the
$|x^2\!-\!y^2\!\!>$ ground state to the $|xy\!\!>$ orbital are
strongly enhanced with respect to the exchange via $|yz\!\!>$ and
$|zx\!\!>$ states. This strong anisotropy in orbital space is
directly translated into a strong anisotropy in spin space via
spin-orbit coupling.  A strong ferromagnetic coupling arises e.g.
due to virtual hopping processes via $|xy,{\rm Cu_L}\!\!>$ into
the $|p_{x,A}\!\!>$ and $|p_{y,B}\!\!>$ ligand states. The hopping
process takes the form of "ring exchange" which avoids unfavorable
doubly occupied sites by involving the excitations along the ring
$\rm Cu_L\rightarrow O_A\rightarrow Cu_R\rightarrow O_B\rightarrow
Cu_L$ \cite{Aharony}. The predominant contribution of the $|xy\!>$
orbital to superexchange causes an easy-axis out-of-plane
anisotropy. In the present geometry this anisotropy, as compared
to a 180$^\circ$ bond, is additionally enhanced due to a strongly
reduced isotropic superexchange between $|x^2\!-\!y^2,{\rm
Cu_L}\!\!>$ and $|x^2\!-\!y^2,{\rm Cu_R}\!\!>$, which may be
reduced furthermore for bond angles slightly larger than
90$^\circ$ because of a cancellation of the leading order ferro-
and antiferromagnetic contributions to the isotropic exchange
\cite{Aharony}.

Our ESR data clearly show that the spin dynamics of the Cu-O
chains of LCCO is governed by a strong anisotropy of the magnetic
interactions in the paramagnetic regime. In this respect LCCO may
not be unique. For instance, in $\rm Li_2CuO_2$ a surprisingly
broad ESR signal (\dh$\sim 4000$ Oe) has been reported in the
submillimeter wavelength range \cite{Ohta}. Complementary to our
result, specific heat measurements of $\rm La_5Ca_9Cu_{24}O_{41}$
in a magnetic field reveal a strong magnetic anisotropy in the AF
{\em ordered} state at low $T$, suggesting even an Ising-like
character of the magnetism of the chains \cite{Ammerahl}.
Moreover, our analysis is in agreement with the recent
observations of a large spin-wave gap in $\rm Li_2CuO_2$
\cite{Boehm} and of a large size of the ordered moment in $\rm
Ca_2Y_2Cu_5O_{10}$ \cite{fong}, which both point towards a strong
easy-axis anisotropy of edge-sharing Cu-O plaquettes.

In summary, a very broad  ESR absorption line
of \cu\ ions in the chains of
\lcco\ single crystals is
observed in the paramagnetic regime. Our analysis reveals that the
line width is two orders of magnitude larger than one expects from
conventional estimates of the anisotropy of the magnetic exchange
interaction. This gives strong experimental evidence for a
significant amplification of the influence of spin-orbit coupling
on magnetic superexchange in edge-sharing Cu-O structures, as
suggested recently by theoretical calculations
\cite{Yushan,Aharony}. The commonly accepted point of view on
copper oxides as good model systems for studies of the isotropic
Heisenberg spin magnetism has thus to be revised for certain
bonding geometries.

We gratefully acknowledge useful discussions with G.A. Sawatzky,
B. Keimer, and E. M\"{u}ller-Hartmann. This work was supported by the
Deutsche Forschungsgemeinschaft through SFB 341.

\newpage

\newpage


\end{document}